\documentclass[a4paper,11pt]{article}
\usepackage{pos}

\def\msbar{{\overline{\rm MS}}}

\title{Form factors in semileptonic decay of D mesons}

\author*[a,b]{Tinghong Shen}
\author[b,c,d]{Ying Chen}
\author[b,c]{Ming Gong}
\author[e]{Dong-Hao Li}
\author[f,g]{\\Keh-Fei Liu}
\author[b,d]{Zhaofeng Liu}
\author[a]{Zhenyu Zhang}

\affiliation[]{\normalsize{\bf \sffamily \hspace{50mm}($\chi$QCD Collaboration)}}

\affiliation[a]{Hubei Nuclear Solid Physics Key Laboratory, School of Physics and Technology, Wuhan University, Wuhan, Hubei 430072, China}

\affiliation[b]{Institute of High Energy Physics, Chinese Academy of Sciences,
Beijing 100049, China}

\affiliation[c]{School of Physical Sciences, University of Chinese Academy of Sciences, Beijing 101408, China}

\affiliation[d]{Center for High Energy Physics, Henan Academy of Sciences, Zhengzhou 450046, China}

\affiliation[e]{MOE Frontiers Science Center for Rare Isotopes and School of Nuclear Science and Technology, Lanzhou University, Lanzhou 730000, China}

\affiliation[f]{Department of Physics and Astronomy, University of Kentucky, Lexington, Kentucky 40506, USA}
\affiliation[g]{
Nuclear Science Division, Lawrence
Berkeley National Laboratory, Berkeley, California 94720, USA}

\emailAdd{liuzf@ihep.ac.cn}
\emailAdd{zhenyuzhang@whu.edu.cn}
\emailAdd{lidonghao@ihep.ac.cn}

\abstract{We study the vector, scalar and tensor form factors for the semileptonic process $D\rightarrow K$ by using lattice Quantum Chromodynamcs (QCD). Chiral lattice fermions are used in our study: overlap fermion for the valence quark and domain-wall fermion for the sea. The 2+1-flavor configurations are from the RBC-UKQCD Collaborations with an inverse lattice spacing $1/a=2.383(9)$ GeV. A modified $z$-expansion taking into account valence quark mass dependence is used to fit our numerical results of the form factors at a fixed pion mass $\sim360$ MeV in the sea. At the physical valence quark mass point we give the preliminary results $f_+(0)=f_0(0)=0.760(39)$ and $f_T(0)=0.733(50)$ with only statistical uncertainties. For $f_T$ the number is given in the $\msbar$ scheme at a scale of 2 GeV.}

\FullConference{The 41st International Symposium on Lattice Field Theory (LATTICE2024)\\
 28 July - 3 August 2024\\
Liverpool, UK\\}

\begin{document}
\maketitle

\section{Introduction}
Semileptonic decays of D mesons, which involve the transition of a charmed meson into a lighter meson, a lepton, and a neutrino, provide opportunities to probe the properties of heavy quarks in a relatively clean environment. They are connected to the Cabibbo-Kobayashi-Maskawa (CKM) matrix, which describes the mixing and transitions between different quark flavors. The semileptonic decay rates depend on the elements of the CKM matrix, particularly $|V_{cd}|$ and $|V_{cs}|$, as well as the hadronic matrix elements, which are parameterized as form factors. Precise calculations of these form factors not only enable a more accurate determination of CKM matrix elements but also provide a test bed for the Standard Model, potentially highlighting discrepancies that could indicate new physics.

The hadronic matrix element involved in $D\rightarrow P(=\pi, K)$ is conventionally parameterized as
\begin{equation}
\langle P(p')|V_\mu|D(p)\rangle=f_+(q^2)\left(p_\mu+p'_\mu-\frac{M_D^2-M_P^2}{q^2}q_\mu\right)+f_0(q^2)\frac{M_D^2-M_P^2}{q^2}q_\mu,
\label{eq:ff}
\end{equation}
where $V_\mu=\bar x\gamma_\mu c$ with $x=d, s$ indicating the daughter light quark and $q=p-p'$. By using the partially conserved vector current relation ($\partial_\mu V_\mu=(m_c-m_x)S$), one has
\begin{equation}
\langle P(p')|S|D(p)\rangle=f_0(q^2)\frac{M_D^2-M_P^2}{m_c-m_x},
\label{eq:f0}
\end{equation}
where $m_{c/x}$ is the charm/light quark mass and $S=\bar x c$. The meson mass $M_D$ or $M_P$ can be obtained from meson two-point functions in Euclidean space on the lattice. By calculating the three-point functions involving the vector/scalar operator and the initial and final state mesons, one can get the hadronic matrix elements and then the form factors.

Besides $f_+$ and $f_0$, we also calculate the form factor $f_T$ for the tensor current $T_{\mu\nu}=\bar x\sigma_{\mu\nu}c$, which is defined by
\begin{equation}
\langle P(p')|T_{\mu\nu}|D(p)\rangle=\frac{2f_T(q^2)}{M_D+M_P}\left[p'_\mu p_\nu-p'_\nu p_\mu\right].
\label{eq:ft}
\end{equation}
This tensor form factor can contribute to flavor-changing neutral current process $c\rightarrow ul^+l^-$ through loop effects in the Standard Model and to tree-level process $D\rightarrow \pi(K)l\nu$ in new physics models.

\section{Lattice setup}
We use the 2+1-flavor configurations from the RBC/UKQCD Collaborations with domain-wall fermions in the sea~\cite{RBC:2010qam}. The parameters of the configurations are given in Table~\ref{tab:confs}. All lattices are of volume $L^3\times T=32^3\times64$. The results reported in this paper are from the ensemble labeled as f006, which has a pion mass around 360 MeV in the sea. The computation of correlation functions and data analyses on the other two ensembles are in progress.
\begin{table}[bht]
    \caption{The 2+1-flavor configurations used in this work. The residual mass of the dynamical fermion $am_{\rm res}$
is in the two-flavor chiral limit from Ref.~\cite{RBC:2010qam}.
$N_{\rm conf}$ is the number of configurations, and $N_{\rm src}$ the number of point sources used or to be used on each configuration.}
    \label{tab:confs}
    \centering
\begin{tabular}{cccccc}
\hline\hline
$a^{-1}$ (GeV) & Label & $am_l^{\rm sea}/am_s^{\rm sea}$ & $L^3\times T$ & $N_{\rm conf}\times N_{\rm src}$ & $am_{\rm res}$ \\
\hline
2.383(9) & \texttt{f004} & 0.004/0.03 & $32^3\times64$ & $628\times1$ & 0.0006664(76) \\
         & \texttt{f006} & 0.006/0.03 & $32^3\times64$ & $42\times32$ & \\
         & \texttt{f008} & 0.008/0.03 & $32^3\times64$ & $49\times16$ & \\
\hline\hline
\end{tabular}
\end{table}

We use Overlap fermions for the valence light, strange and charm quarks as we did when calculating the decay constants of charmed mesons in Ref.~\cite{Li:2024vtx}.
The massless overlap Dirac operator~\cite{Neuberger:1997fp} is defined as
\begin{equation}
D_{\rm ov}(\rho)=1 + \gamma_5 \varepsilon (\gamma_5 D_{\rm w}(\rho)),
\end{equation}
where $\varepsilon$ is the matrix sign function and $D_{\rm w}(\rho)$ is the usual Wilson fermion operator,
except for a negative mass parameter $- \rho = 1/2\kappa -4$ with $\kappa_c < \kappa < 0.25$ and $\kappa_c$ corresponding to a massless Wilson operator. In practice, we set $\kappa = 0.2$ corresponding to $\rho = 1.5$. The massive overlap Dirac operator is defined as
\begin{eqnarray}
D_m &=& \rho D_{\rm ov} (\rho) + m\, (1 - \frac{D_{\rm ov} (\rho)}{2}) \nonumber\\
       &=& \rho + \frac{m}{2} + (\rho - \frac{m}{2})\, \gamma_5\, \varepsilon (\gamma_5 D_{\rm w}(\rho)).
\end{eqnarray}
To accommodate the SU(3) chiral symmetry, it is convenient to use the chirally regulated field
$\hat{\psi} = (1 - \frac{1}{2} D_{\rm ov}) \psi$ instead of $\psi$ in the interpolation operators and the currents.
This is equivalent to leaving the currents unmodified and instead adopting the effective propagator
\begin{equation}
G \equiv D_{\rm eff}^{-1} \equiv (1 - \frac{D_{\rm ov}}{2}) D^{-1}_m = \frac{1}{D_c + m},
\end{equation}
where $D_c = \frac{\rho D_{\rm ov}}{1 - D_{\rm ov}/2}$ satisfies $\{\gamma_5, D_c\}=0$~\cite{Chiu:1998gp}.

The bare valence quark masses used in this study are listed in Table~\ref{tab:val} in lattice units. The lightest pion mass is around 220 MeV. Thus, an extrapolation is needed to arrive the physical light quark mass.
\begin{table}[tbh]
    \caption{ Valence quark masses used in this study.
The physical mass point of valence charm quark is close to 0.492 in lattice units.}
    \centering
\begin{tabular}{ll}
\hline\hline
   $am_l$ & 0.00460, 0.00765, 0.01290 0.02400 \\
   $m_\pi$ & $\sim$ 220 $-$ 500 MeV  \\
   $am_s$ &   0.037, 0.040, 0.043, 0.046, 0.049, 0.052 \\
   $am_c$  & 0.450, 0.492, 0.500, 0.550   \\
\hline\hline
\end{tabular}
\label{tab:val}
\end{table}
For the strange or charm quark the physical point is in the middle of our quark masses.

To increase statistics we use 32 sources on each of the 42 configurations of ensemble f006.
For each measurement a point source quark propagator for each flavor is computed to construct the two-point functions for the $D$ and $K$ mesons. Then the light quark propagator is used to construct a sequential source for calculating a sequential propagator for the charm quark. The three-point function is constructed from the sequential propagator and the strange quark propagator.

The initial $D$ meson is set to be at rest. The 3-momentum at the current can be changed freely and the 3-momentum of the daughter meson is fixed by momentum conservation.

That is to say, we calculate the following two-point function
\begin{equation}
C_{K/D}(t,\vec p)= \sum\limits_{\vec{x}} \langle 0 | \mathcal{O}_{K/D}(\vec{x},x_0)\mathcal{O}_{K/D}^{\dagger}(\vec s,s_0)|0\rangle e^{-i\vec p\cdot (\vec x-\vec s)},
\label{eq:lqcd-2pt}
\end{equation}
where $t\equiv x_0-s_0$ is the time displacement in lattice units between the source point $s$ and sink point $x$, and three-point function
\begin{eqnarray}
C_3(\tau,T_s)=\sum_{\vec x,\vec y}\langle0|\mathcal{O}_K(\vec x,x_0)J(\vec y, y_0)\mathcal{O}_D^\dagger(\vec s, s_0)|0\rangle e^{-i\vec p'\cdot (\vec x-\vec s)}e^{-i\vec q\cdot (\vec y-\vec s)},
\end{eqnarray}
where $\tau=y_0-s_0$ and $T_s=x_0-s_0$. The interpolation operators are $\mathcal{O}_K=\bar l\gamma_5 s$ and $\mathcal{O}_D=\bar l\gamma_5 c$. The current $J$ is $V_\mu=\bar x \gamma_\mu c$ or $S=\bar x c$.

The separation $T_s$ in the time direction between the initial hadron and the final hadron is taken to be the maximal value on our lattice: $T_s=T/2=32$. $T_s$
is varied to check if it is big enough so that the ground states dominate in the three-point functions.

Thanks to the good chiral property of overlap fermions we can use Eq.~(\ref{eq:f0}) to determine the form factor $f_0$ without renormalization constants since $Z_sZ_m=1$.
The local vector current on the lattice needs a finite renormalization constant $Z_V$ due to discretization effects. For our lattice setup this constant was calculated in Ref.~\cite{Bi:2023pnf} as well as the renormalization constant for the tensor current. The numbers are
\begin{equation}
    Z_V=Z_A=1.0789(10),\quad\quad\quad Z_T^\msbar(2\mbox{ GeV})=1.157(11).
\end{equation}

\section{Data analyses and preliminary results}
\begin{figure}[thb]
\centering
\includegraphics[width=0.49\columnwidth]{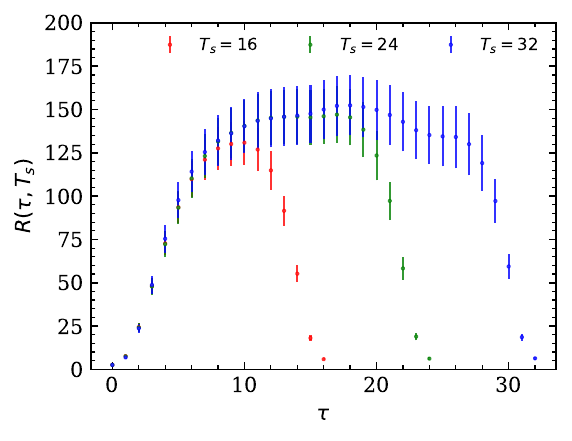}
\includegraphics[width=0.49\columnwidth]{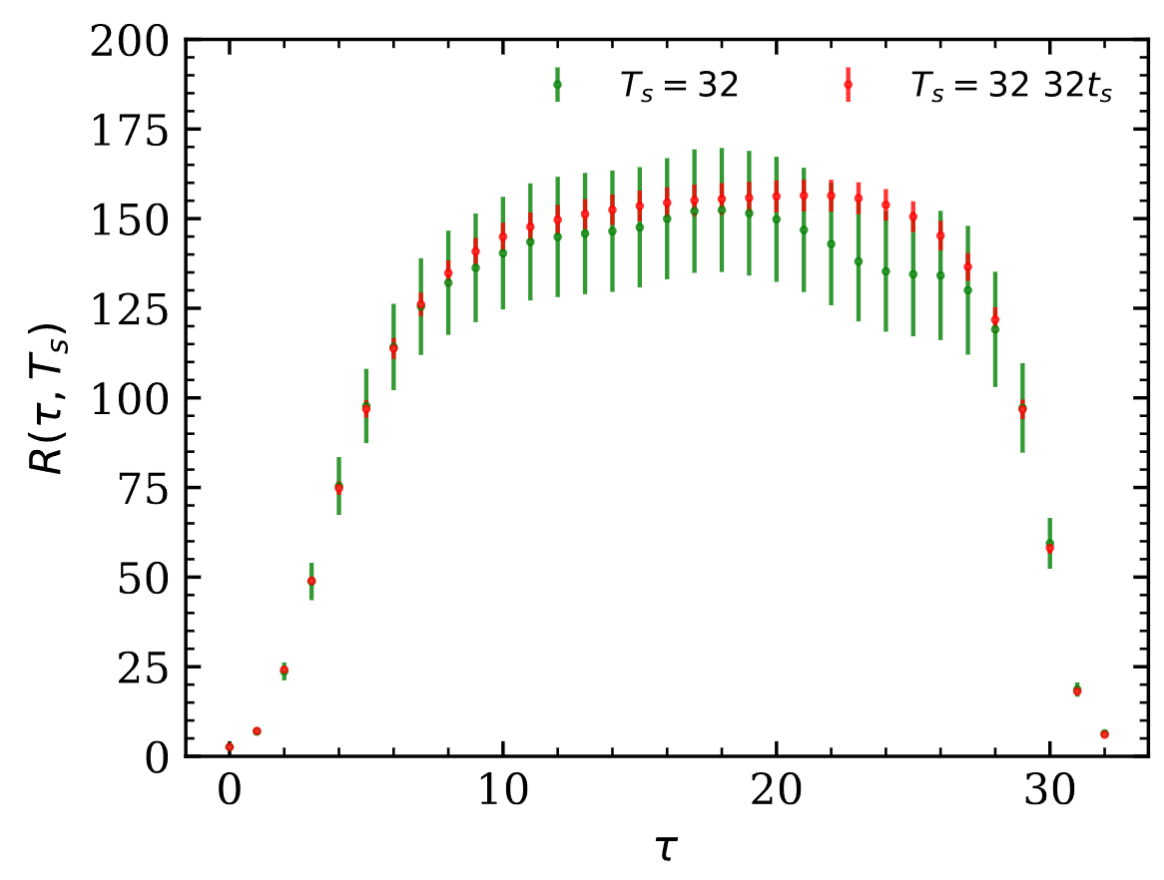}
\caption{Left: The ratio $R(\tau,T_s)$ in Eq.~(\ref{eq:R}) as a function of $\tau$ for the scalar current with $T_s=16, 24$ or $32$ at valence quark masses $am_l=0.024$, $am_s=0.037$ and $am_c=0.450$. Right: A better plateau is obtained as the number of sources increases from 1 (green) to 32 (red) for $T_s=32$.}
\label{fig:R}
\end{figure}

A ratio of three and two-point functions is constructed as
\begin{equation}
    R(\tau,T_s)=\frac{C_3(\tau,T_s)}{C_K(T_s-\tau)C_D(\tau)},
    \label{eq:R}
\end{equation}
for the vector or the scalar current.
In the large time separation limit ($\tau\rightarrow\infty,(T_s-\tau)\rightarrow\infty$), the ground state contribution dominates in both two- and three-point functions such that the above ratio $R(\tau,T_s)$ approaches a plateau independent of $\tau$ and $T_s$. We plot $R(\tau,T_s)$ in the left panel of Fig.~\ref{fig:R} as a function of $\tau$ for the scalar current with $T_s=16, 24$ or $32$ at valence quark masses $am_l=0.02400$, $am_s=0.03700$ and $am_c=0.450$. Here both the $D$ and $K$ mesons are at rest. Only one source is used on each configuration for this check.
As we can see, $T_s=16$ is too small for the ground states to saturate the correlation functions. While for the two cases $T_s=24$ and $32$, $R(\tau,T_s)$ reaches a common plateau around $\tau\sim T_s/2$. The right panel of Fig.~\ref{fig:R} shows that a much better plateau is obtained as we increase the number of sources from 1 (green) to 32 (red).
In the following calculation we always set $T_s=32$, which is the maximal value available on our lattice.

We perform simultaneous fits to two- and three-point functions. The fitting functions are
\begin{eqnarray}
    &&C_{K/D}(t,\vec p)=A_{K/D}\left(e^{-E_{K/D}t}+e^{-E_{K/D}(64-t)}\right),\\
    &&C_3(t)=A_3\left(e^{-E_D t} e^{-E_K(32-t)}\pm e^{-E_D(64-t)}e^{-E_K(t-32)}\right).
\end{eqnarray}
Both the two- and three-point functions are (anti-)symmetric around $t=T/2=32$. Thus, we fold the data as we fit in a range satisfying $t\le 32$. The fitting range is chosen by requiring $\chi^2/{\rm dof}\le 1.2$ and the steadiness of the fitted parameters.

\begin{figure}[thb]
\centering
\includegraphics[width=0.5 \columnwidth]{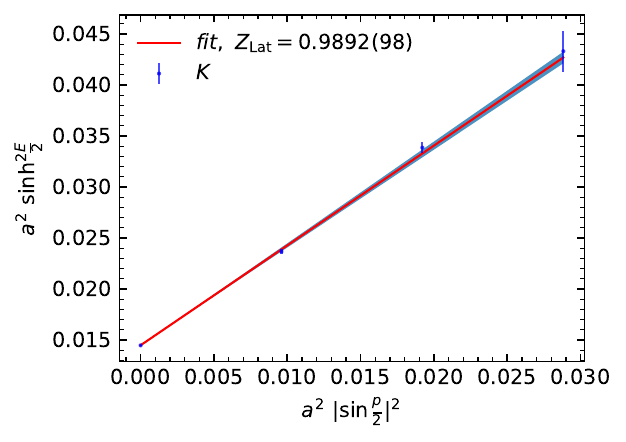}
\caption{The Dispersion relation for $K$ meson at valence quark masses $am_l=0.024$ and $am_s=0.037$.}
\label{fig:dispersion-K}
\end{figure}

We use four values for the 3-momentum of the daughter $K$ meson (in lattice units):
\begin{equation}
\vec p'=\frac{2\pi}{L}\{(0,0,0),(0,0,1),(0,1,1),(1,1,1)\}.
\end{equation}
For each momentum case, the data with symmetric momentum modes related by rotation are averaged to increase statistics.
The dispersion relation for the $K$ meson is checked in Fig.~\ref{fig:dispersion-K}.
Here we use the discrete dispersion relation to fit the energies of the $K$ meson:
\begin{equation}
\sinh^2\frac{E_K}{2} = \sinh^2\frac{M_K}{2} + Z_\mathrm{Lat}^2 \cdot \bigg|\sin^2\frac{\vec p^\prime}{2}\bigg|^2
\end{equation}
A nice linear behavior is observed in Fig.~\ref{fig:dispersion-K} and the effective speed of light from the fit is consistent with one: $Z_\mathrm{Lat}=0.9892(98)$.

\begin{figure}[thb]
\centering
\includegraphics[width=0.5 \columnwidth]{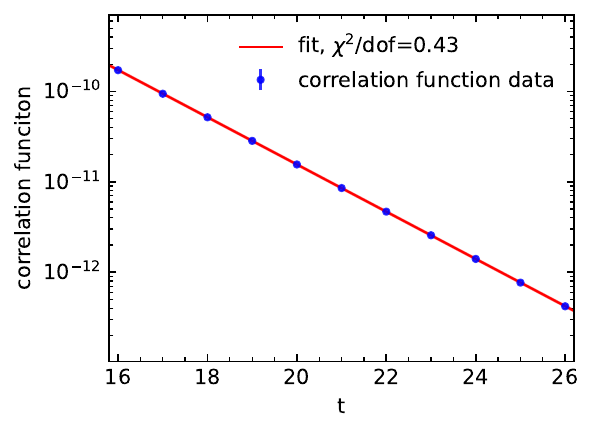}
\caption{The scalar current data, obtained with quark masses $am_l=0.024$, $am_s=0.037$, and $am_c=0.450$, are compared with the fitting results.}
\label{fit_3pt}
\end{figure}

A comparison of three-point correlators and the fitted function from the simultaneous fit is plotted in Fig.~\ref{fit_3pt}.
Here the fitting range is $t\in[19, 23]$ and the momentum for the $K$ meson is $\vec p'=(2\pi/L)(0,0,0)$.

\begin{figure}[thb]
\centering
\includegraphics[width=0.48 \columnwidth]{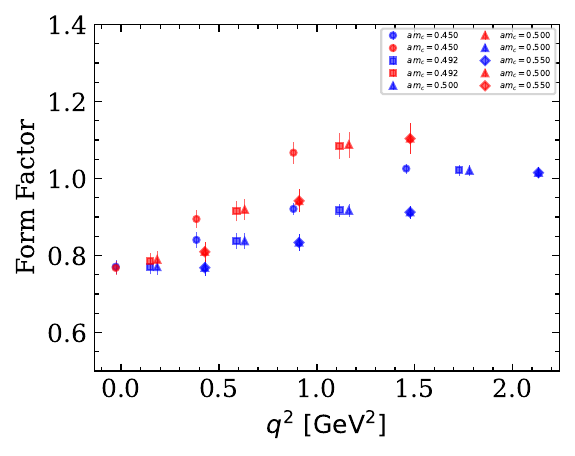}
\includegraphics[width=0.48 \columnwidth]{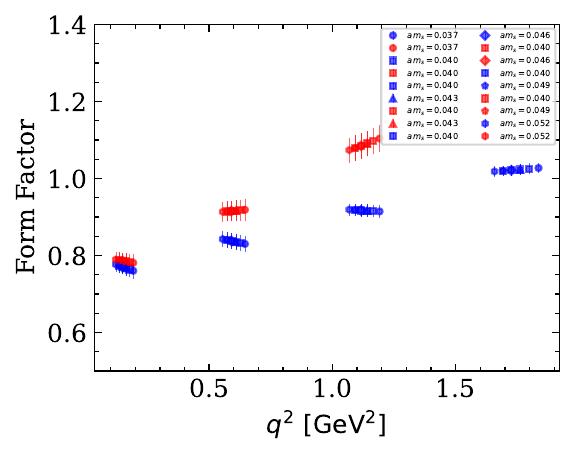}
\caption{Dependence of form factors $f_+$ (red) and $f_0$ (blue) on charm (left panel) and strange (right panel) quark masses.}
\label{fig:mass-dep}
\end{figure}

From the simultaneous fits one can get the form factors $f_+$ and $f_0$ from the fitting parameters $A_{K/D}$, $A_3$ and $E_{K/D}$ at three/four $q^2$ values. We get consistent $f_0$ from the three-point functions of the vector current and the scalar density after taking $Z_V$ into account. The dependence of the form factors on charm or strange quark mass is shown in Fig.~\ref{fig:mass-dep}, where the errors are from Jackknife analyses.
On the left panel, we fix the light and strange quark masses to $am_l=0.024$, $am_s=0.046$ and change the charm quark mass.
While on the right panel, we fix the light and charm quark masses to $am_l=0.024$ and $am_c=0.492$. Similarly one obtains the form factor $f_T$.

To obtain the shape of form factors in the whole physical range of $q^2$ at physical valence quark masses, a modified $z$-expansion is used to fit our lattice results of the form factors at all quark masses. The mapping between $q^2$ and $z$ is
\begin{equation}
    z(q^2;t_0)=\frac{\sqrt{t_+-q^2}-\sqrt{t_+-t_0}}{\sqrt{t_+-q^2}+\sqrt{t_+-t_0}},
\end{equation}
where $t_+=(M_D+M_K)^2$ and $t_0$ can be chosen to minimize the maximum value of $|z|$ in the whole range of $q^2$. For our $D\rightarrow K$ process it is convenient to set $t_0=0$ so that $z=0$ corresponds to $q^2=0$.
The $z$-expansion functions are
\begin{eqnarray}
    f_+(q^2;m_c,m_s,m_l)&=&\frac{1}{1-q^2/m_{D_s^*}^2}\sum_{i=0}^n a_i D_i z^i,\\
    f_0(q^2;m_c,m_s,m_l)&=&\frac{1}{1-q^2/m_{D_{s0}^*}^2}\sum_{i=0}^n b_i D_i z^i,\\
    f_T(q^2;m_c,m_s,m_l)&=&\frac{1}{1-q^2/m_{D_s^*}^2}\sum_{i=0}^n a_i^T D_i z^i,
\end{eqnarray}
where
\begin{eqnarray}
    D_i=1+c_{i1}\left(m_\pi^2-(m_\pi^{\rm phys})^2\right)+c_{i2}\left(m_{\eta_s}^2-(m_{\eta_s}^{\rm phys})^2\right)+c_{i3}\left(m_{J/\psi}-m_{J/\psi}^{\rm phys}\right)
\end{eqnarray}
taking into account the valence quark mass dependence. Form factors $f_+$ and $f_0$ are fitted together with the condition $f_+(0)=f_0(0)$ at $q^2=0$, which requires $a_0=b_0$. The meson masses $m_\pi$, $m_{\eta_s}$ (fictitious), $m_{J/\psi}$ and the pole mass $m_{D_s^*}$ at our unphysical quark masses are taken from our previous work done on the same gauge ensembles for meson decay constants~\cite{Li:2024vtx}. The pole mass $m_{D_{s0}^*}$ is obtained by fitting newly calculated two-point functions on the ensemble f006. The physical mass point in isospin-symmetric QCD is determined by~\cite{ParticleDataGroup:2024cfk,Borsanyi:2020mff,Bazavov:2017lyh}
\begin{equation}
    m_\pi^{\rm phys}=134.98\mbox{ MeV},\quad\quad m_{\eta_s}^{\rm phys}=689.89\mbox{ MeV},\quad\quad m_{J/\psi}^{\rm phys}=3.0969\mbox{ GeV}.
\end{equation}

We tried $n=0,1,2$ and $3$ for the truncation in the $z$-expansion. The fittings were also repeated as we fix the pole masses to their PDG values~\cite{ParticleDataGroup:2024cfk}. The center values and uncertainties of the form factors at $q^2=0$ stabilize once $n\ge1$. We choose $n=1$ to give our preliminary results.

\begin{figure}[thb]
    \centering
    \includegraphics[width=0.48\linewidth]{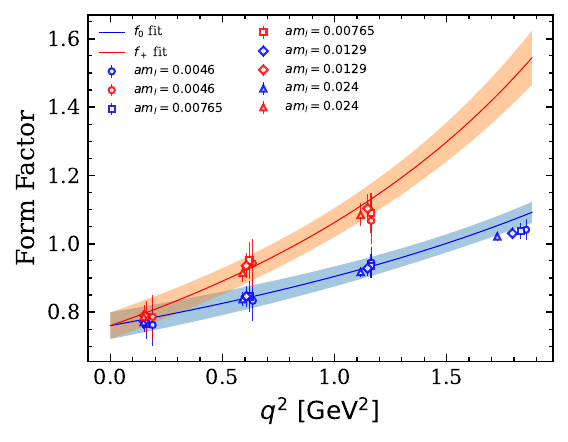}
    \includegraphics[width=0.48\linewidth]{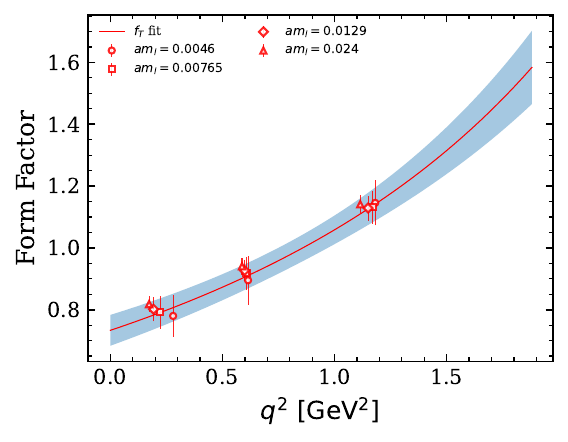}
    \caption{Left: The form factors \( f_0 \) and \( f_+ \) with \( am_s = 0.046 \) and \( am_c = 0.492 \) are compared with the $z$-expansion fitting results with $n=1$. The blue curve represents the fitting result for \( f_0 \), while the red curve corresponds to the fitting result for \( f_+ \). Right: The form factor \( f_T \), calculated with \( am_s = 0.046 \) and \( am_c = 0.492 \), is compared with the $z$-expansion fitting results with $n=1$. In both graphs the statistical and fitting error is indicated by a band.}
    \label{zexp_f0f+}
\end{figure}

Finally, the form factors \( f_0 \), \( f_+ \), and \( f_T \), evaluated at the physical valence quark masses, are shown as functions of \( q^2 \) in Fig.~\ref{zexp_f0f+}. For clarity in this figure we only plot part of the data points with \( am_s = 0.046 \) and $am_c=0.492$.  At $q^2=0$ we obtain $f_+(0)=f_0(0)=0.760(39)$ and $f_T(0)=0.733(50)$, where the errors are statistical only. Note our sea quark mass is not at the physical point yet.

\section{Summary}
We calculate the form factors for the semileptonic decay $D\rightarrow K$ with overlap valence quark on 2+1-flavor domain-wall fermion configurations. Preliminary results are presented in this paper at one lattice spacing with an unphysical pion mass $\sim360$ MeV in the sea. At physical valence quark masses we find $f_+(0)=f_0(0)=0.760(39)$ and $f_T(0)=0.733(50)$, where the errors come from statistics and the $z$-expansion fitting. Data generation and analyses at other light sea quark masses are underway so that we can extrapolate the results to the physical sea quark mass point in the future.

\section*{Acknowledgements}
We thank the RBC-UKQCD collaborations for sharing the domain wall fermion configurations.
This work is supported in part by National Key Research and Development Program of China under Contract No. 2023YFA1606002 and by the National Natural Science Foundation of China (NSFC) under Grants No. 12075253, No. 11935017, No. 12192264, No. 12293060, No. 12293063, No. 12293065, No. 12447154 and 12070131001 (CRC 110 by DFG and NNSFC) and Joint Fund of Research utilizing Large-Scale Scientific Facility of the NSFC and CAS under Contract No. U2032114.
K. L. is partially supported by the U.S. DOE Grant No. DE-SC0013065 and DOE, Office of Nuclear Physics under the umbrella of the Quark-Gluon Tomography (QGT) Topical Collaboration, with Award No. DE-SC0023646.
The GWU code~\cite{Alexandru:2011ee,Alexandru:2011sc} is acknowledged.
The computations were performed on the HPC clusters at Institute of High Energy Physics (Beijing).

\end{document}